\documentclass[aps,showpacs,showkeys,pre,twocolumn,superscriptaddress,floatfix]{revtex4}
\usepackage{graphicx,bm,amssymb,amsmath,dcolumn,hyperref}
\usepackage{algorithmic}
\usepackage{multirow}
\usepackage{enumitem}
\usepackage{color}
\usepackage{amsfonts}
\usepackage{amsthm}

\definecolor{blue}{rgb}{0,0,1}
\definecolor{red}{rgb}{1,0,0}
\definecolor{green}{rgb}{0,1,0}

\begin{document}
\newtheorem{lemma}{Lemma}
\newtheorem{algorithm}{Algorithm}
\newcommand{\be}{\begin{equation}}
\newcommand{\ee}{\end{equation}}
\newcommand{\bes}{\begin{eqnarray}}
\newcommand{\ees}{\end{eqnarray}}
\newcommand{\symdif}{\triangle}
\newcommand{\ind}{\mathbf{1}}
\renewcommand{\Pr}{\mathbb{P}}
\newcommand{\ZZ}{\mathbb{Z}}
\newcommand{\EE}{\mathbb{E}}
\newcommand{\LL}{\mathbb{L}}
\newcommand{\fL}{{\mathfrak L}}
\newcommand{\scrA}{{\mathcal A}}
\newcommand{\scrB}{{\mathcal B}}
\newcommand{\scrL}{{\mathcal L}}
\newcommand{\scrN}{{\mathcal N}}
\newcommand{\scrS}{{\mathcal S}} 
\newcommand{\scrs}{{\mathcal s}}
\newcommand{\scrP}{{\mathcal P}}
\newcommand{\scrM}{{\mathcal M}}
\newcommand{\scrO}{{\mathcal O}}
\newcommand{\scrR}{{\mathcal R}}
\newcommand{\scrC}{{\mathcal C}}
\newcommand{\scrl}{{\mathcal l}}
\newcommand{\dm}{d_{\rm min}}
\newcommand{\rhojunction}{\rho_{\rm j}}
\newcommand{\rhojunctionLim}{\rho_{{\rm j},0}}
\newcommand{\rhobranch}{\rho_{\rm b}}
\newcommand{\rhobranchLim}{\rho_{{\rm b},0}}
\newcommand{\rhononbridge}{\rho_{\rm n}}
\newcommand{\rhononbridgeLim}{\rho_{{\rm n},0}}
\newcommand{\leafFreeHull}{H_{\rm lf}}
\newcommand{\bridgeFreeHull}{H_{\rm bf}}
\newcommand{\percolationCluster}{C_1}
\newcommand{\leafFreeCluster}{C_{\rm lf}}
\newcommand{\bridgeFreeCluster}{C_{\rm bf}}
\newcommand{\dF}{d_{\rm F}}
\newcommand{\dB}{d_{\rm B}}
\newcommand{\dR}{d_{\rm R}}
\newcommand{\dH}{d_{\rm H}}
\newcommand{\dE}{d_{\rm E}}
\newcommand{\ontop}[2]{\genfrac{}{}{0pt}{}{#1}{#2}}
\newcommand{\vc}{v_{\mathrm{c}}}

\newcommand{\verweiFig}{\includegraphics[scale=0.7]{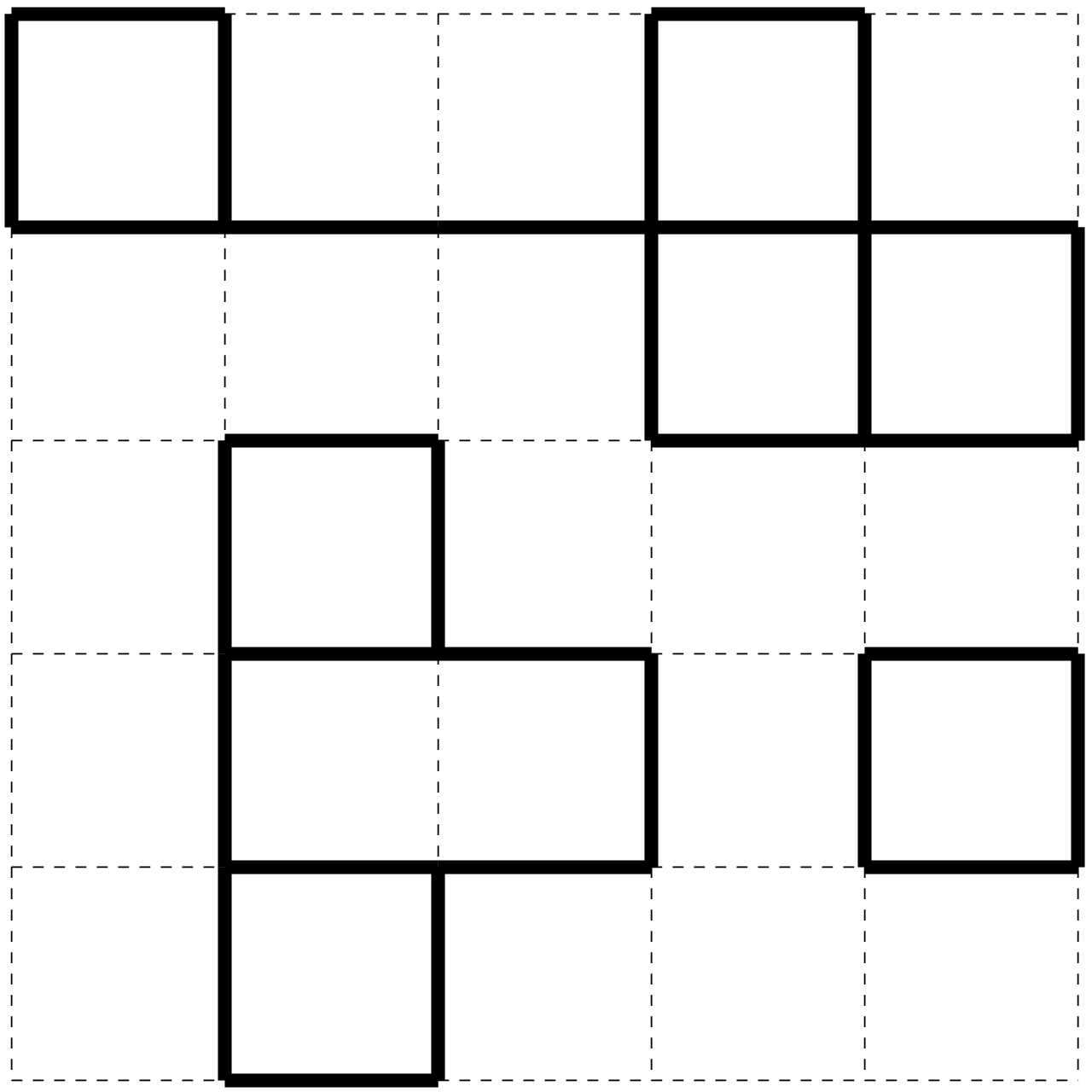}}
\newcommand{\configurationFig}{\includegraphics[trim = 0mm 0mm 0mm 0mm, clip, scale=0.68]{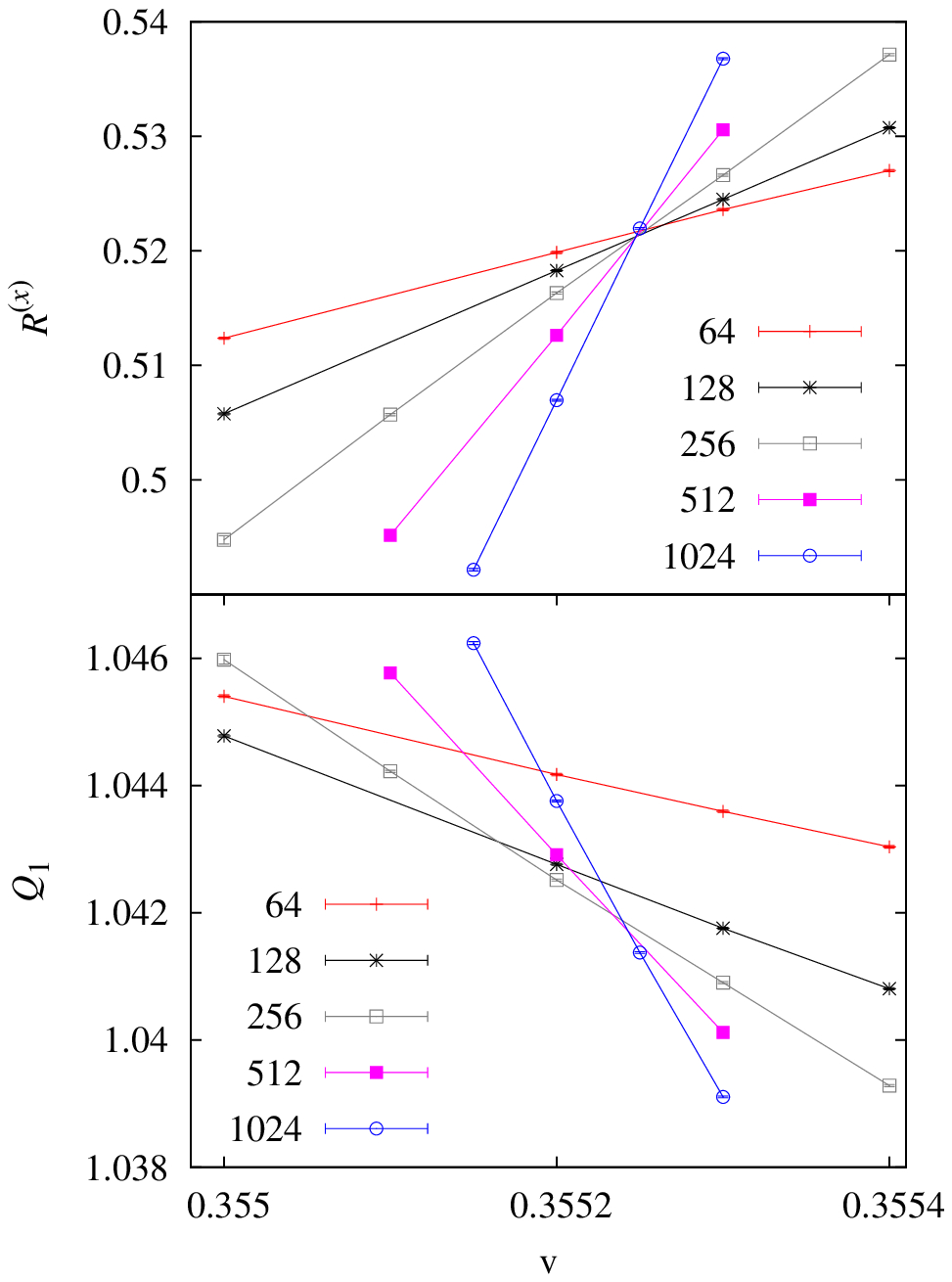}}
\newcommand{\densitiesFig}{\includegraphics[scale=0.7]{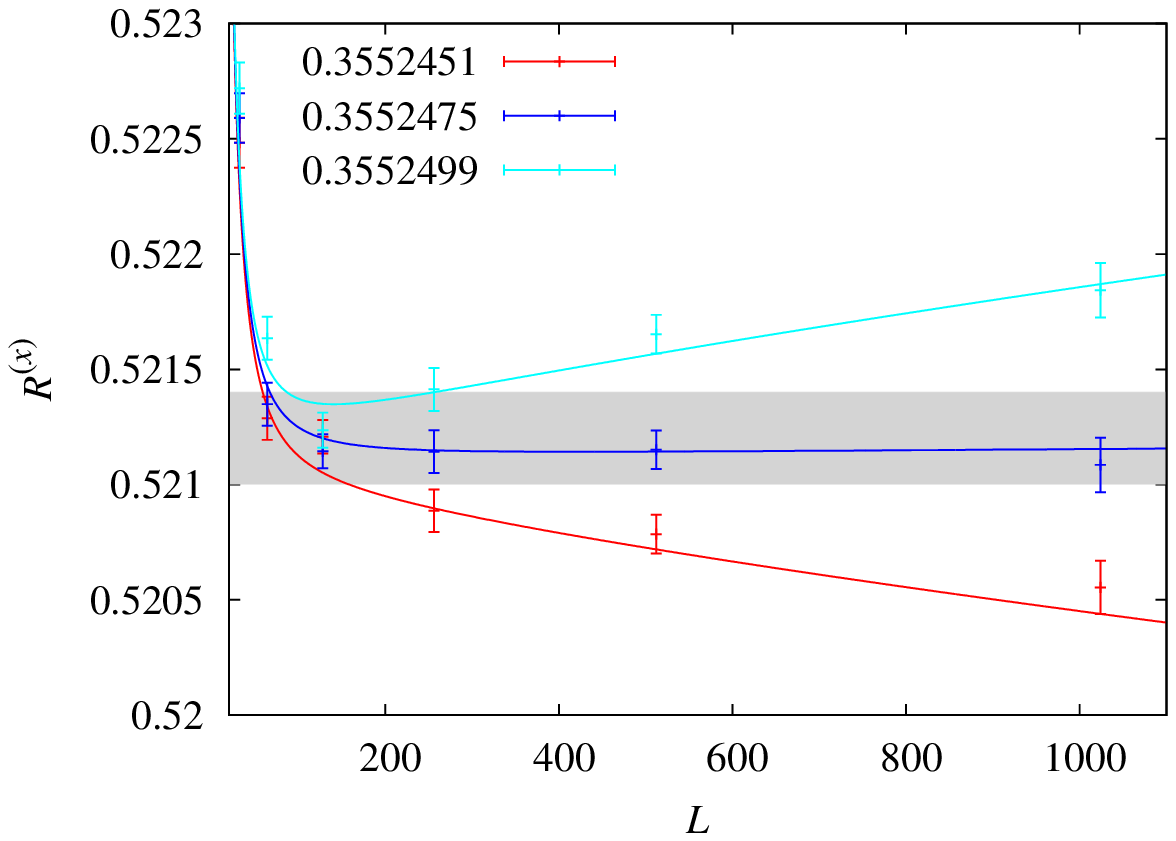}}
\newcommand{\leafFreeFig}{\includegraphics[scale=0.68]{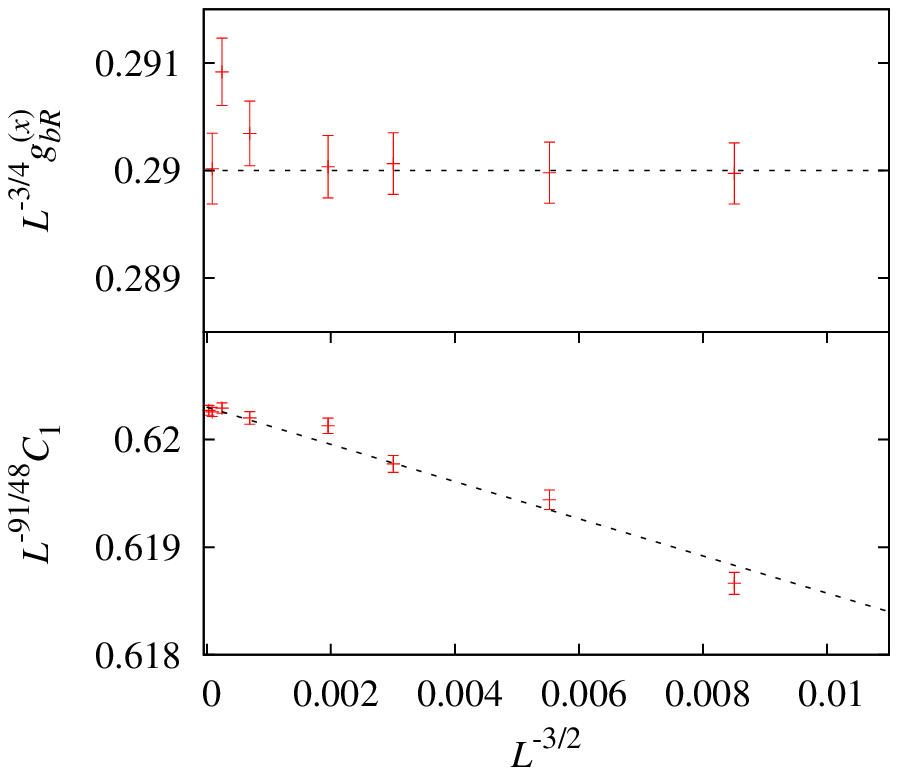}}
\newcommand{\backboneFig}{\includegraphics[scale=0.68]{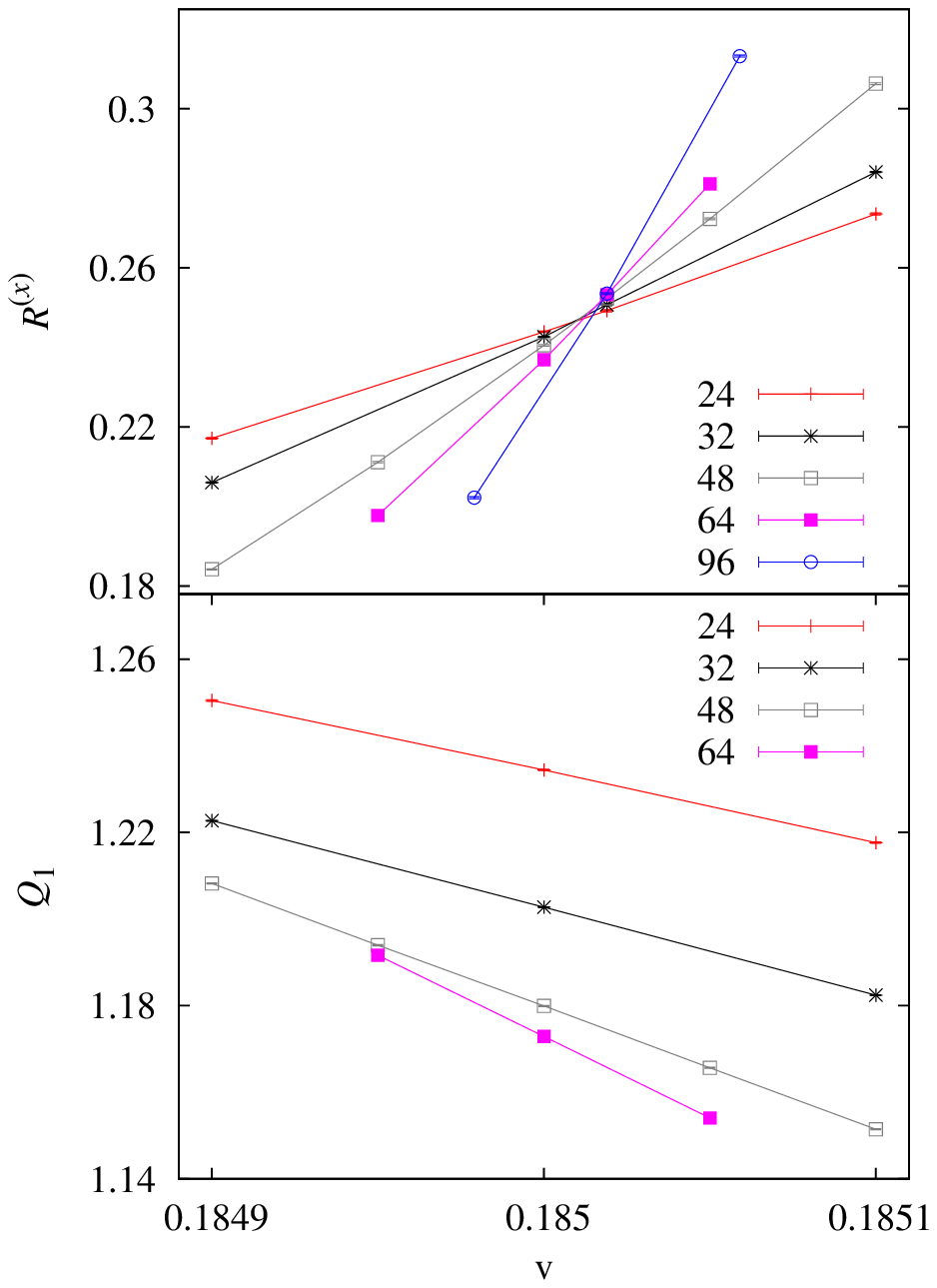}}
\newcommand{\hullFig}{\includegraphics[scale=0.68]{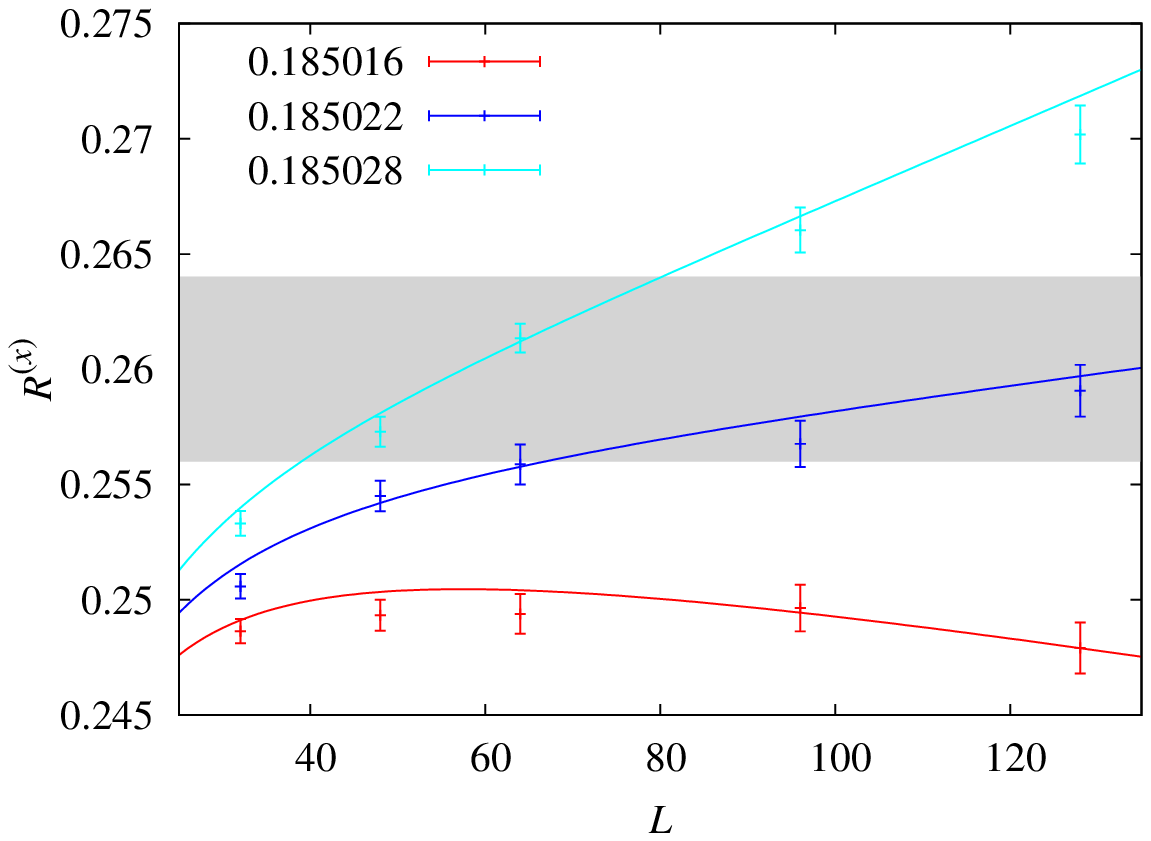}}
\newcommand{\loopFigA}{\includegraphics[trim = 35mm 190mm 40mm 30mm, clip, scale=0.325]{fig7a.eps}}
\newcommand{\loopFigB}{\includegraphics[trim = 35mm 190mm 40mm 30mm, clip, scale=0.325]{fig7b.eps}}

\title{Leaf-excluded percolation in two and three dimensions}
\date{\today}
\author{Zongzheng Zhou}
\affiliation{School of Mathematical Sciences, Monash University, Clayton, Victoria~3800, Australia}
\author{Xiao Xu}
\affiliation{Hefei National Laboratory for Physical Sciences at Microscale and Department of Modern Physics, University of Science and
  Technology of China, Hefei, Anhui 230026, China}
\author{Timothy M. Garoni}
\email{tim.garoni@monash.edu}
\affiliation{School of Mathematical Sciences, Monash University, Clayton, Victoria~3800, Australia}
\author{Youjin Deng}
\email{yjdeng@ustc.edu.cn}
\affiliation{Hefei National Laboratory for Physical Sciences at Microscale and Department of Modern Physics, University of Science and
  Technology of China, Hefei, Anhui 230026, China}

\begin{abstract}
We introduce the \emph{leaf-excluded} percolation model, which corresponds to independent bond percolation conditioned on the absence of
leaves (vertices of degree one). We study the leaf-excluded model on the square and simple-cubic lattices via Monte Carlo simulation,
using a worm-like algorithm. By studying wrapping probabilities, we precisely estimate the critical thresholds to be $0.355\,247\,5(8)$
(square) and $0.185\,022(3)$ (simple-cubic).  Our estimates for the thermal and magnetic exponents are consistent with those for
percolation, implying that the phase transition of the leaf-excluded model belongs to the standard percolation universality class.

\end{abstract}
\maketitle

\section{Introduction}
\label{Introduction}
Graphical models, i.e. statistical-mechanical models in which the configuration space consists of certain bond configurations drawn on a
lattice, play a fundamental role in the theory of critical phenomena. Examples include percolation, and more generally the Fortuin-Kasteleyn
random-cluster model, as well as dimers and various loop models. In the latter two cases, the models are in fact examples of
``forbidden-degree'' models, in which only bond configurations which preclude specified vertex degrees are allowed; for dimers, all
degrees higher than 1 are forbidden, while loop models forbid all odd degrees. 

\begin{figure}
  \includegraphics[scale=0.4]{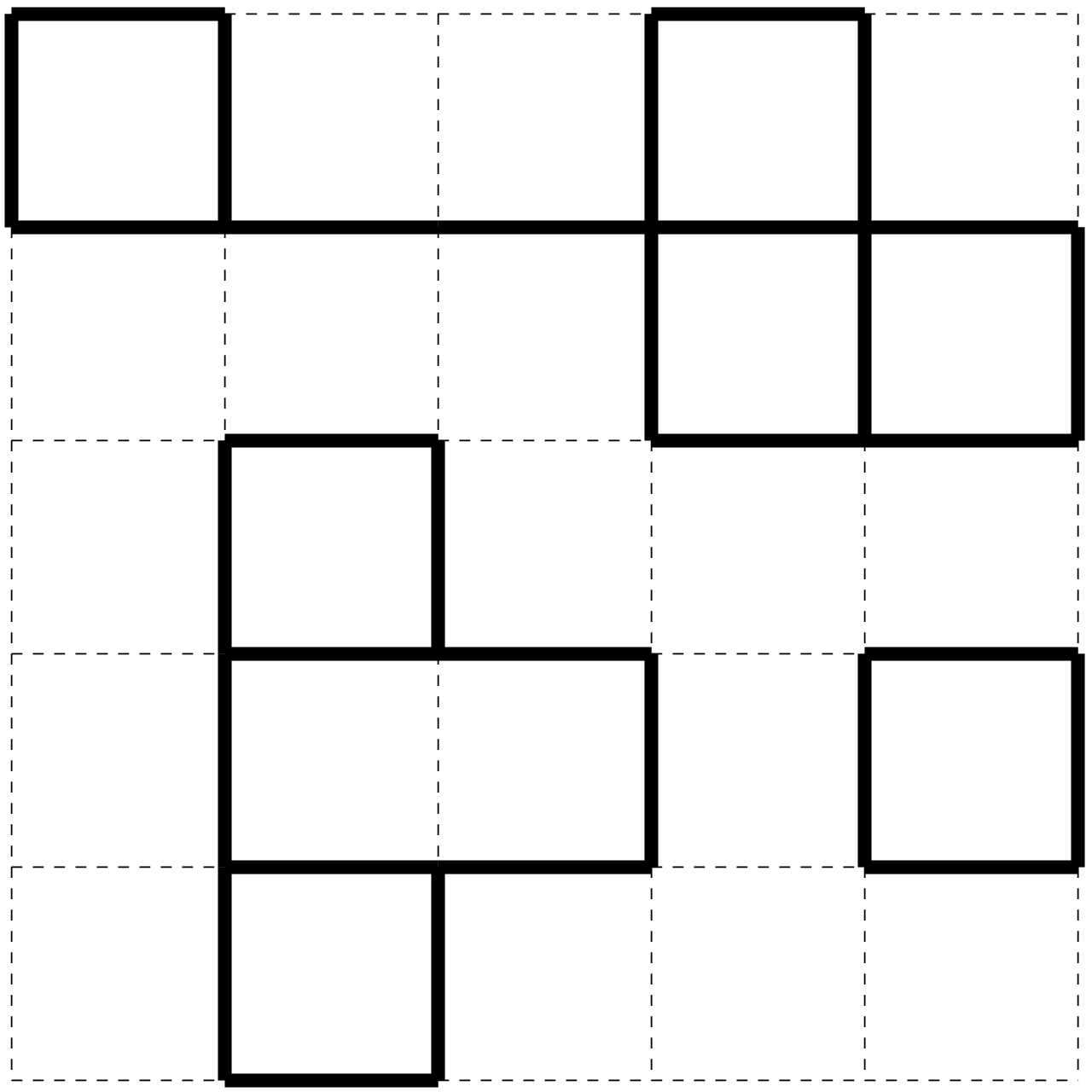}
  \caption{A typical configuration (denoted by bold edges) of the leaf-excluded model on a $6\times6$ patch of the square lattice.}
 \label{Fig:Config}
 \end{figure}

In this article, we introduce and study another example of a forbidden-degree model, the {\em leaf-excluded} model, which forbids bond
configurations containing vertices of degree 1 (i.e. {\em leaves}). Consider a finite connected graph $G=(V,E)$, and let
\begin{equation}
\label{leaf-excluded configuration space}
\Omega = \{A\subseteq E : \ d_A(i)\neq 1 \},
\end{equation}
where $d_A(i)$ denotes the degree of vertex $i$ in the spanning subgraph $(V,A)$. As an example, 
Fig.~\ref{Fig:Config} illustrates a typical element of $\Omega$ in the case where $G$ is a $6\times6$ patch of the square lattice. 
In this case, $\Omega$ is the set of all ways of drawing bond configurations such that each site has degree 0, 2, 3 or 4.

The leaf-excluded model on $G$ chooses random configurations $A\in\Omega$ according to the distribution 
\begin{equation}
\label{leaf-excluded distribution}
\Pr(A) \propto v^{|A|},
\end{equation}
where $v>0$ is a (temperature-like) bond fugacity, and $|A|$ denotes the number of bonds in the configuration $A$.

The model defined by~\eqref{leaf-excluded configuration space} and~\eqref{leaf-excluded distribution} is equivalent to considering standard
independent bond percolation and conditioning on the absence of leaves. This conditioning then introduces non-trivial correlations between
the edges. Note that, on the square lattice, if we additionally forbid degree 3 vertices, the resulting model coincides with the high
temperature (and low temperature) expansion of the Ising model.  On the square lattice therefore, the definition of the leaf-excluded model
lies precisely half way between the definitions of standard percolation (no vertex degrees forbidden) and the Ising loop representation
(both leaves and degree 3 vertices forbidden). It is therefore natural to ask to which universality class does the leaf-excluded model
belong?

One of the main goals of percolation theory in recent decades has been to understand the geometric structure of percolation clusters,
following the pioneering work of Stanley~\cite{Stanley77}. 
Recently, the present authors~\cite{XuWangZhouGaroniDeng14} studied the geometric structure of percolation clusters by classifying the
bridges present in clusters into two types: branches and junctions. A bridge was defined to be a branch if and only if at least one of the
two clusters produced by its deletion is a tree. It was found that the leaf-free clusters, obtained by deleting the branches from
percolation clusters, have the same fractal dimension and hull dimension as the original percolation clusters.

The set of all such leaf-free configurations coincides with $\Omega$ as defined in~\eqref{leaf-excluded configuration space}. We emphasize,
however, that in~\cite{XuWangZhouGaroniDeng14} these configurations were generated by applying a \emph{burning
  algorithm}~\cite{HerrmannHongStanley84} to standard bond percolation configurations, whereas in the current work they are sampled directly
from the distribution~\eqref{leaf-excluded distribution}. The probability distribution on these configurations studied in~\cite{XuWangZhouGaroniDeng14} is
therefore very different to the distribution that we consider here. Nevertheless, based on the observations
from~\cite{XuWangZhouGaroniDeng14}, one might expect that the leaf-excluded model should belong to the percolation universality class. In
this article, we present a careful numerical study which confirms this picture.

Due to the non-trivial combinatorial constraint inherent in the definition of $\Omega$, to efficiently generate random samples from the
leaf-excluded model requires a suitable Markov-chain Monte Carlo algorithm. We introduce a worm-like algorithm for this purpose.  Using this
algorithm, we simulate the leaf-excluded model on the square and simple-cubic lattices with periodic boundary conditions.  We estimate the
critical threshold $\vc$ by studying the finite-size scaling of wrapping probabilities. Wrapping probabilities are believed to be universal,
and have been successfully applied to the estimation of critical thresholds of several
models~\cite{WangZhouZhangGaroniDeng13,NewmanZiff00,XuWangLvDeng14}. By simulating precisely at our estimated $\vc$, we then estimate the
thermal exponent $y_t = 1/\nu$ and magnetic exponent $y_h=d - \beta/\nu$. Here the exponent $\beta$ describes the critical scaling of the
percolation probability $P_\infty\sim (v-\vc)^\beta$, while $\nu$ describes that of the correlation length $\xi\sim |v-\vc|^{-\nu}$. Our
results for critical exponents and universal amplitudes strongly suggest that the phase transition of the leaf-excluded model belongs to the
standard percolation universality class. 

The remainder of this paper is organized as follows. Sec.~\ref{Algorithm Quantities} introduces the worm-like algorithm and the
observables measured in our simulations.  Numerical results are summarized and analyzed in Section~\ref{Results}.  A brief discussion is
then given in Section~\ref{Discussion}.

\section{Algorithm and Observables}
\label{Algorithm Quantities}

\subsection{Monte Carlo algorithm}
\label{Worm algorithm}
In this section we describe a Markov-chain Monte Carlo algorithm for simulating the leaf-excluded model, which is similar in spirit to a
worm algorithm~\cite{ProkofevSvistunov01}. Worm algorithms provide very effective tools for simulating models on configuration spaces
which are subject to non-trivial combinatorial constraints. The key idea underlying worm algorithms is to first enlarge the configuration
space by including ``defects", and to then move these defects via random walk. Numerical studies have shown that worm algorithms typically
provide highly efficient Monte Carlo methods~\cite{DengGaroniSokal07_worm,Wolff09a,Wolff09b,Wolff10a,Wolff10b}.

To simulate the leaf-excluded model, we therefore consider an enlarged configuration space in which up to two leaves are permitted.
For clarity, it is convenient to define the algorithm on an arbitrary (finite and connected) graph $G=(V,E)$.
The space of worm configurations is then
$$
\scrS = \{(A,u,v)\in E\times V^2 : d_A(i) \neq 1 \text{ for } i\neq u,v\}.
$$
The algorithm proceeds as follows. Let $\symdif$ denote symmetric difference of sets. At each time step, we perform precisely one of the
following three possible updates, chosen at random with respective probabilities $p_1,p_2,p_3$:
\begin{enumerate}
\item Set $(A,u,v)\mapsto(A,v,u)$
\item Choose uniformly random $w\in V$ and set $(A,u,v)\mapsto(A,w,v)$ if $d_A(u)\neq 1$ and $d_A(w)\neq 1$.
\item Do the following:
  \begin{enumerate}
  \item\label{choose neighbour} Choose uniformly random $w\sim u$ 
  \item Propose $(A,u,v)\mapsto(A\symdif uw,w,v)$, and
    accept with probability $\min[1,v^{|A\symdif uw|-|A|}]$, provided $(A\symdif uw,w,v)\in\scrS$.
  \end{enumerate}
\end{enumerate}
After each update, if the new state is leaf-excluded, we measure observables.
In our simulations, we used $p_1=p_2=1/4$ and $p_3=1/2$.

We note that, unlike the case of the worm algorithm for the Ising model~\cite{ProkofevSvistunov01}, there is no particular reason for using
two defects in our algorithm, and in fact the above algorithm can be easily modified to use any fixed number of defects; including one
defect. We also note that it would be somewhat of a misnomer to refer to the above algorithm as a \emph{worm} algorithm; for a state
$(A,u,v)\in\scrS$, it will not be true in general that $u$ and $v$ are connected by occupied bonds, and so in general there is no
\emph{worm} as such. This is in contrast to worm algorithms for Eulerian subgraphs (e.g. Ising high temperature graphs),
where the handshaking lemma demands that the two defects be connected.

\subsection{Sampled quantities}
\label{Measured quantities}
We simulated the leaf-excluded model on the $L\times L$ square lattice for system sizes up to $L=1024$, and on the $L\times L\times L$ simple-cubic
lattice for system sizes up to $L=96$. For each system size, approximately $10^8$ samples were produced.

For each sampled leaf-excluded bond configuration, we measured the following observables.
 \begin{enumerate}
 \item The number of occupied bonds $\scrN_b$.
 \item The size of the largest cluster $\scrC_1$.
 \item The cluster-size moments $\scrS_m  = \sum_{\scrC}|\scrC|^m$ with $m=2,4$,
   where the sum is over all clusters $\scrC$.
 \item The indicators $\scrR^{(x)}$, $\scrR^{(y)}$, $\scrR^{(z)}$ for the event that a cluster wraps around the lattice in the $x$, $y$, or
   $z$ direction, respectively.
 \end{enumerate}

From these observables, we calculated the following quantities:
 \begin{enumerate}
 \item The mean size of the largest cluster $C_1 = \langle \scrC_1\rangle$, which scales as $\sim L^{y_h}$ at the critical point
   $\vc$.
 \item The mean size of the cluster at the origin, $\chi = \langle \scrS_2\rangle/L^d$, which at $\vc$ scales as $\sim L^{2y_h - d}$.
 \item The dimensionless ratios
   \begin{equation}
     Q_1  = \frac{\langle {\scrC_1}^2\rangle}{\langle \scrC_1\rangle^2}\;,\;\;\; 
     Q_2  = \frac{\langle 3{\scrS_2}^2 - 2\scrS_4\rangle}{\langle {\scrS_2}^2\rangle}\;.
     \label{eq:R}
   \end{equation}

\item The probability that a winding exists in the $x$ direction $R^{(x)} = \langle \scrR^{(x)}\rangle$.  In two dimensions, we also
  measured $R^{(2)}=\langle\scrR^{(x)}\scrR^{(y)}\rangle$, and in three dimensions measured
  $R^{(3)}=\langle\scrR^{(x)}\scrR^{(y)}\scrR^{(z)}\rangle$. $R^{(d)}$ gives the probability that windings simultaneously exist in all $d$
  possible directions.
 \item The covariance of $\scrR^{(x)}$ and $\scrN_b$
\begin{equation}
 \begin{aligned}
   g^{(x)}_{bR} &= \langle \scrR^{(x)} \scrN_b\rangle - \langle \scrR^{(x)}\rangle \langle \scrN_b \rangle,
 \end{aligned}
 \label{eq:g}
 \end{equation}
 which is expected to scale as $\sim L^{y_t}$ at the critical point.
  \end{enumerate}
 
\section{Results}
\label{Results}
\subsection{Fitting methodology}
\label{FittingMethology}
We began by estimating the critical point $\vc$ by performing a finite-size scaling analysis of the ratios $Q_1$, $Q_2$ and wrapping
probabilities $R^{(x)}$, $R^{(d)}$. The MC data for these quantities were fitted to the ansatz
 \be
 \scrO(\epsilon,L) = \scrO_c + \sum_{k=1}^{2}{q_k \epsilon^kL^{ky_t}} + b_1L^{y_i} + b_2L^{y_2}\;,
 \label{Fit:fitnearpc}
 \ee
where $\epsilon=\vc-v$, $y_i$ and $y_2$ are respectively the leading and sub-leading correction exponents, and
${\scrO_c=\scrO(\epsilon=0,L\rightarrow+\infty)}$ is a universal constant. The parameters $q_k,b_1,b_2$ are non-universal amplitudes.
 
We then performed extensive simulations at our best estimate of $\vc$, in order to estimate the critical exponents $y_t$ and $y_h$. These
exponents were obtained by fitting $g^{(x)}_{bR}$, $C_1$ and $\chi$ to the ansatz
 \be
 \scrO(L) = L^{y_\scrO}(a_0 + b_1L^{y_i} + b_2L^{y_2})\;,
 \label{Fit:fitatpc}
 \ee
 where $y_\scrO$ equals $y_t$ for $g^{(x)}_{bR}$, $y_h$ for $C_1$ and $2y_h - d$ for $\chi$, and $a_0$ is a non-universal constant. In all
 fits reported below, we fixed $y_2=-2$, which corresponds to the exact value of the sub-leading correction exponent~\cite{Nienhuis84} for
 percolation.
 
 As a precaution against correction-to-scaling terms that we failed to include in the fit ansatz, we imposed a lower cutoff $L\geq L_{\rm
   min}$ on the data points admitted in the fit, and we systematically studied the effect on the $\chi^2$ value of increasing $L_{\rm min}$.
 Generally, the preferred fit for any given ansatz corresponds to the smallest $L_{\min}$ for which the goodness of fit is reasonable and
 for which subsequent increases in $L_{\min}$ do not cause the $\chi^2$ value to drop by vastly more than one unit per degree of freedom.
 In practice, by ``reasonable'' we mean that $\chi^2/\mathrm{DF}\lessapprox 1$, where DF is the number of degrees of freedom.

 We analyze the data on the square lattice in Sec.~\ref{2Daroundpc} and Sec.~\ref{2Datpc}. The results on the simple cubic lattice are shown
 in Sec.~\ref{3Dresults}.

\subsection{\texorpdfstring{Square lattice near $\vc$}{Square lattice near criticality}}
\label{2Daroundpc}

\begin{figure}
  \includegraphics[scale=0.65]{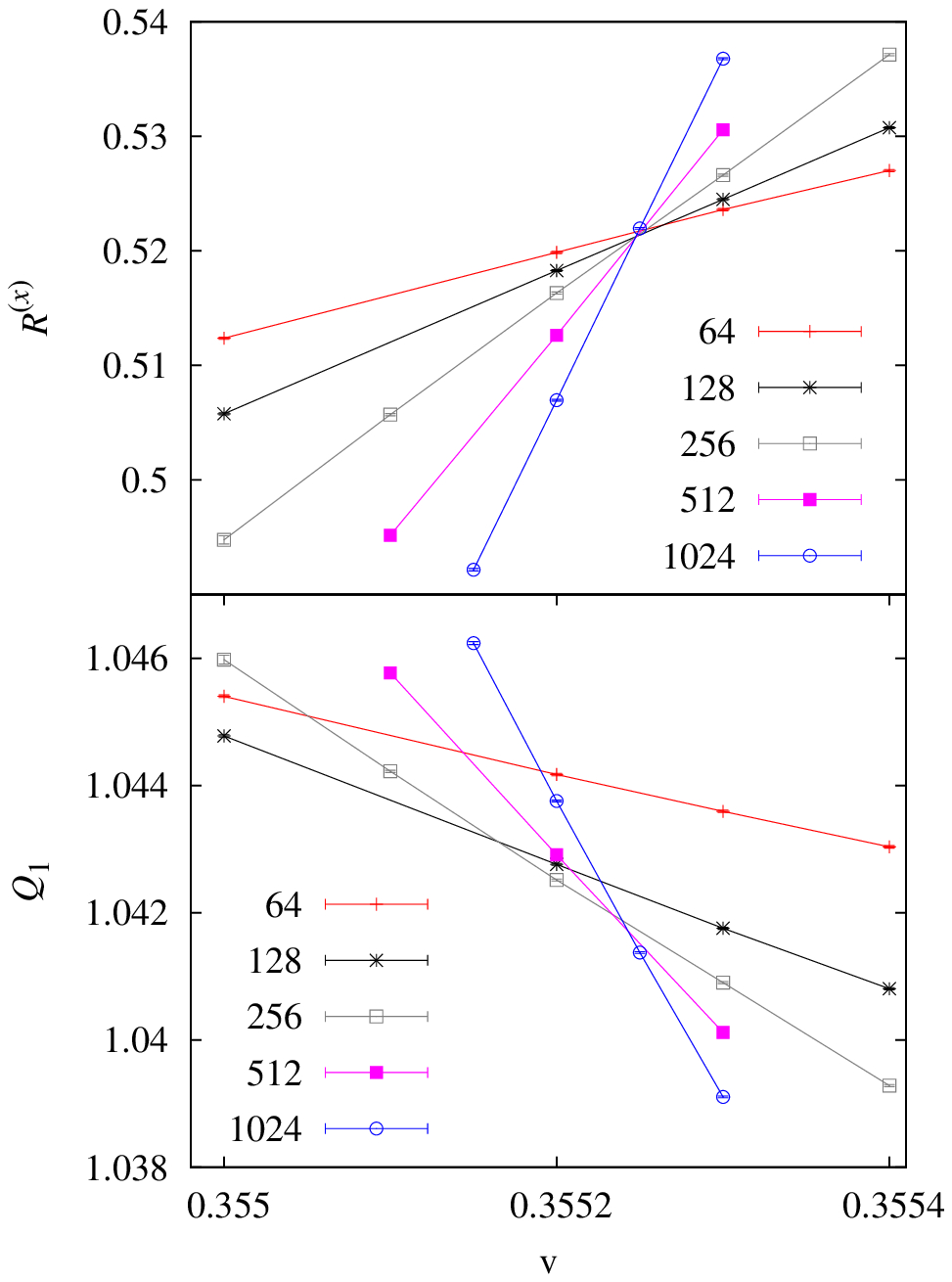}
  \caption{Plots of $R^{(x)}$ and $Q_1$ versus $v$ for the leaf-excluded model on the square lattice.}  
 \label{Fig:RxQ1}
  \end{figure}

 We first study the critical behavior of $R^{(x)}$, $R^{(d)}$ and $Q_1$, $Q_2$ near $\vc$. Fig.~\ref{Fig:RxQ1} plots $R^{(x)}$ and
 $Q_1$ versus $v$. Clearly, $R^{(x)}$ suffers from only very weak corrections to scaling.
 
 We begin by considering $R^{(x)}$. Setting $b_2=0$ and leaving $y_i$ free, we were unable to obtain a stable estimate of $y_i$. The fits
 with two correction terms included (fixing $y_i = -1$) show that $b_1$ is consistent with zero and $b_2=2(1)$ for $L_{\rm min} = 64$.  In
 fact, the data for $R^{(x)}$ with $L_{\rm min} = 128$ can be well fitted even with fixed $b_1=b_2=0$.  We also perform fits with only one of
 $b_1L^{-1}$ or $b_2L^{-2}$ included.  Comparing the various fit results, we estimate $\vc = 0.355\,247\,5(5)$ and $y_t = 0.752(3)$. The
 latter is clearly consistent with $3/4$ for two-dimensional percolation. We also estimate the universal amplitude $R^{(x)}_c =
 0.521\,2(2)$, consistent with the exact value $0.521\,058\,290$~\cite{Pinson94,ZiffLorenzKleban99}.
 
 The fits of $R^{(d)}$ show that it suffers even weaker finite-size corrections.  The amplitudes $b_1$ and $b_2$ are both consistent with
 zero for $L_{\rm min}= 16$.  Again, we also perform fits in which we include only one of these corrections, and also fits in which we
 include neither.  We then estimate $\vc = 0.355\,247\,4(5)$, $y_t = 0.754(3)$ and $R^{(d)}_c = 0.351\,7(1)$, the latter of
 which is consistent with the exact value $0.351\,642\,855$ for standard percolation~\cite{Pinson94,ZiffLorenzKleban99}.
 
 Finally, we fit the data for $Q_1$ and $Q_2$.
 The fits predict a leading correction exponent $y_i=-1.57(5)$ and $-1.7(1)$ respectively.
 We note that this is consistent with the exact value $-3/2$~\cite{Ziff11} for two-dimensional percolation.
 We estimate $\vc = 0.355\,247\,5(5)$ from $Q_1$ and $\vc = 0.355\,247\,5(8)$ from $Q_2$.
 Both of their fits produce $y_t = 0.751(3)$.
 We also estimate the universal amplitudes $Q_{1,c} = 1.041\,47(5)$ and $Q_{2,c} = 1.148\,6(2)$, both of which are consistent with the estimates for standard percolation~\cite{HuBloteDeng12}.

 \begin{figure}
  \includegraphics[scale=0.65]{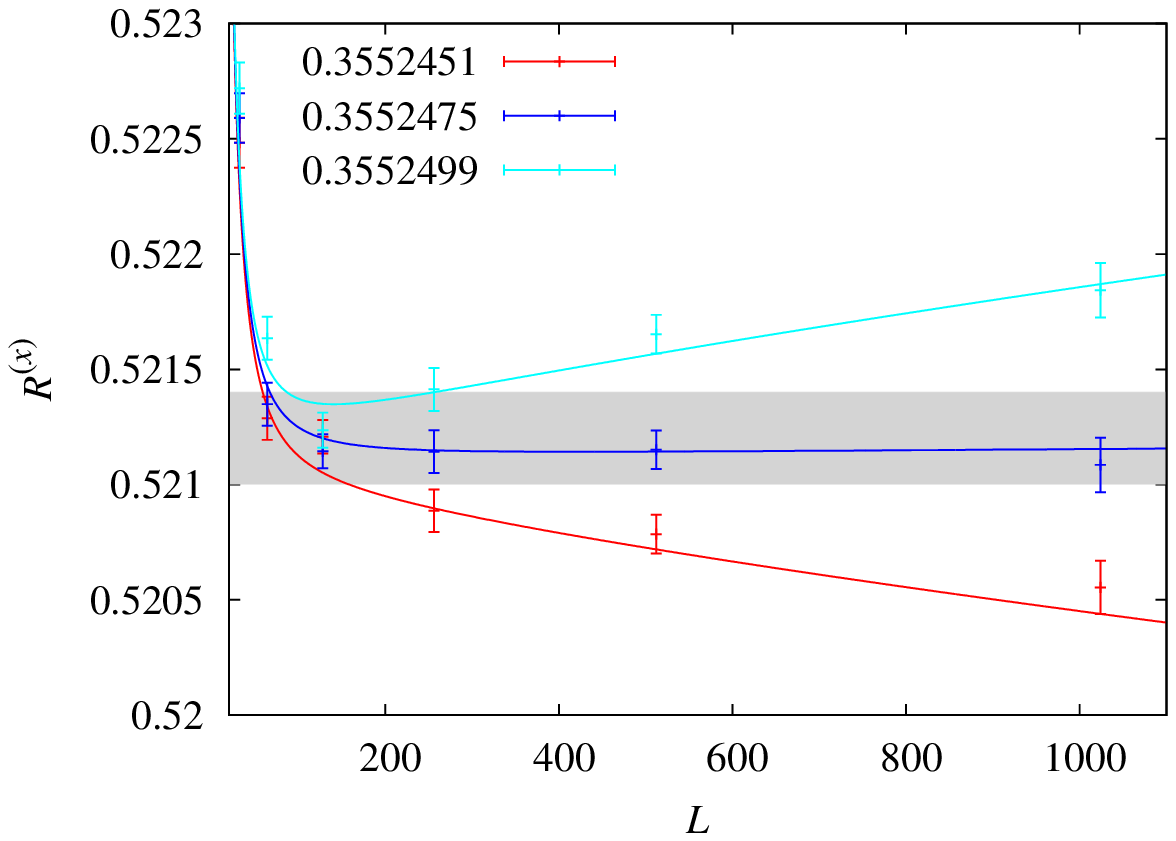}
  \caption{Plots of $R^{(x)}(v,L)$ versus $L$ for fixed values of $v$, for the two-dimensional leaf-excluded model. The curves correspond
    to our preferred fit of the Monte Carlo data. The shaded grey strips indicate an interval of one error bar above and below the estimate
    $R^{(x)}_c=0.5212(2)$.}
 \label{Fig:check_pc_2d}
  \end{figure} 
 
 Our estimates for $\vc$, $y_t$ and the universal wrapping probabilities are summarized in Tab.~\ref{Tab:Results}, where we also report the
 known results for standard percolation. The results strongly suggest that the phase transition of the leaf-excluded model belongs to the
 standard percolation universality class.

In Fig.~\ref{Fig:check_pc_2d}, we illustrate the accuracy of our estimate of $\vc$ by plotting $R^{(x)}$ versus $L$ with $v$ set to our
central estimate of $\vc$, $v=0.355\,247\,5$, and also with $v$ chosen three error bars above and below this estimate.  Precisely at
$v=\vc$, as $L\to\infty$ the data should tend to a horizontal line, whereas the data with $v\neq\vc$ should bend upward or downward.
Fig.~\ref{Fig:check_pc_2d} provides confirmation that the true value of $v_{\mathrm{c}}$ does indeed lie in the interval
$(0.3552451,0.3552499)$. Moreover, the asymptotic flatness of the $R^{(x)}$ curve at our reported central estimate of $v_{\mathrm{c}}$
strongly suggests that our estimate lies very close indeed to the true value of $\vc$.

\subsection{\texorpdfstring{Square lattice at $\vc$}{Square lattice at criticality}}
\label{2Datpc}
To obtain final estimates of $y_t$ and $y_h$, we performed high-precision simulations at a single value of $v$ corresponding to our
estimated threshold $\vc=0.355\,247\,5$, and fitted the data for $g_{bR}^{(x)}$, $C_1$ and $\chi$ to~\eqref{Fit:fitatpc}. The leading correction
exponent was set to $y_i = -3/2$.
 
The fits of $g^{(x)}_{bR}$ show that both the amplitudes $b_1$ and $b_2$ are consistent with zero.  The data for $g^{(x)}_{bR}$ can be well
fitted ($\chi^2/DF<1$ for $L_{\rm min}=24$) even without any corrections.  From the fits, we estimate $y_t = 0.750(1)$,
which is consistent with the estimate in Sec.~\ref{2Daroundpc} but with improved precision.
  
\begin{figure}
  \includegraphics[scale=0.8]{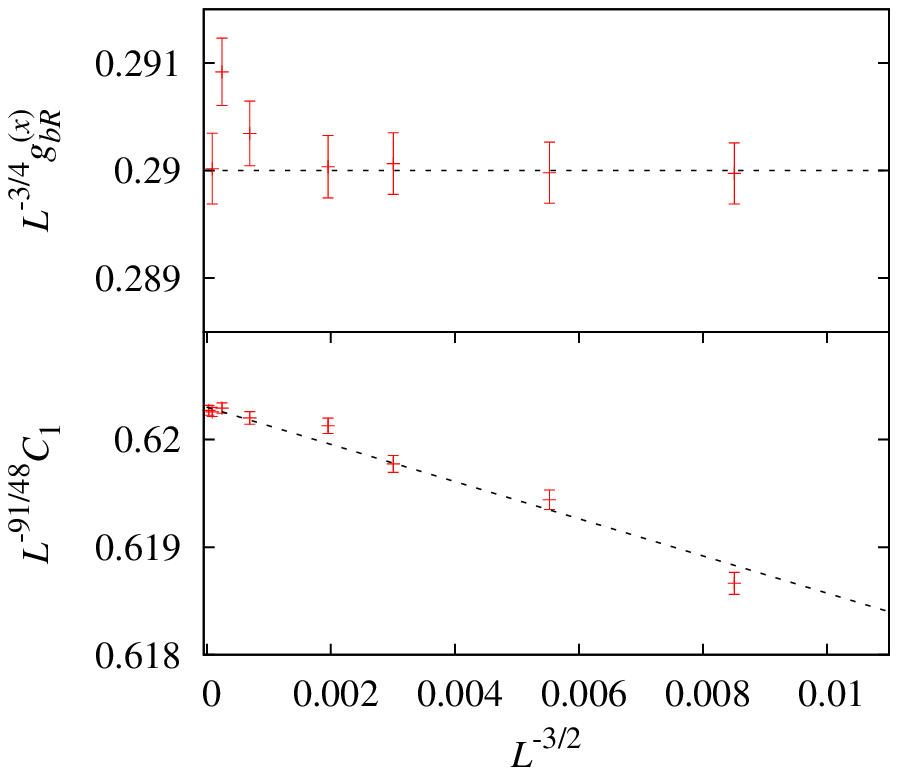}
  \caption{Plots of $L^{-3/4}g^{(x)}_{bR}$ and $L^{-91/48}C_1$ versus $L^{-3/2}$. The straight lines are simply to guide the eye.}
  \label{Fig:2Dytyh}
\end{figure}

The fits of $C_1$ and $\chi$ show a non-zero $b_1$ if only the leading correction term is included in the fits. For comparison, we also
performed fits including only the $b_2L^{-2}$ term, and including both corrections. Both of these fits suggest $y_h = 1.895\,8(1)$, which is
fully consistent with the exact result $y_h = 91/48$ for two-dimensional percolation.
As further illustration, Fig.~\ref{Fig:2Dytyh} shows a plot of $L^{-3/4}g^{(x)}_{bR}$ and $L^{-91/48}C_1$ versus $L^{-3/2}$.
 
\subsection{Simple-cubic lattice}
\label{3Dresults}
We performed an analogous study of the leaf-excluded model on the simple-cubic lattice.

\begin{figure}
  \includegraphics[scale=0.65]{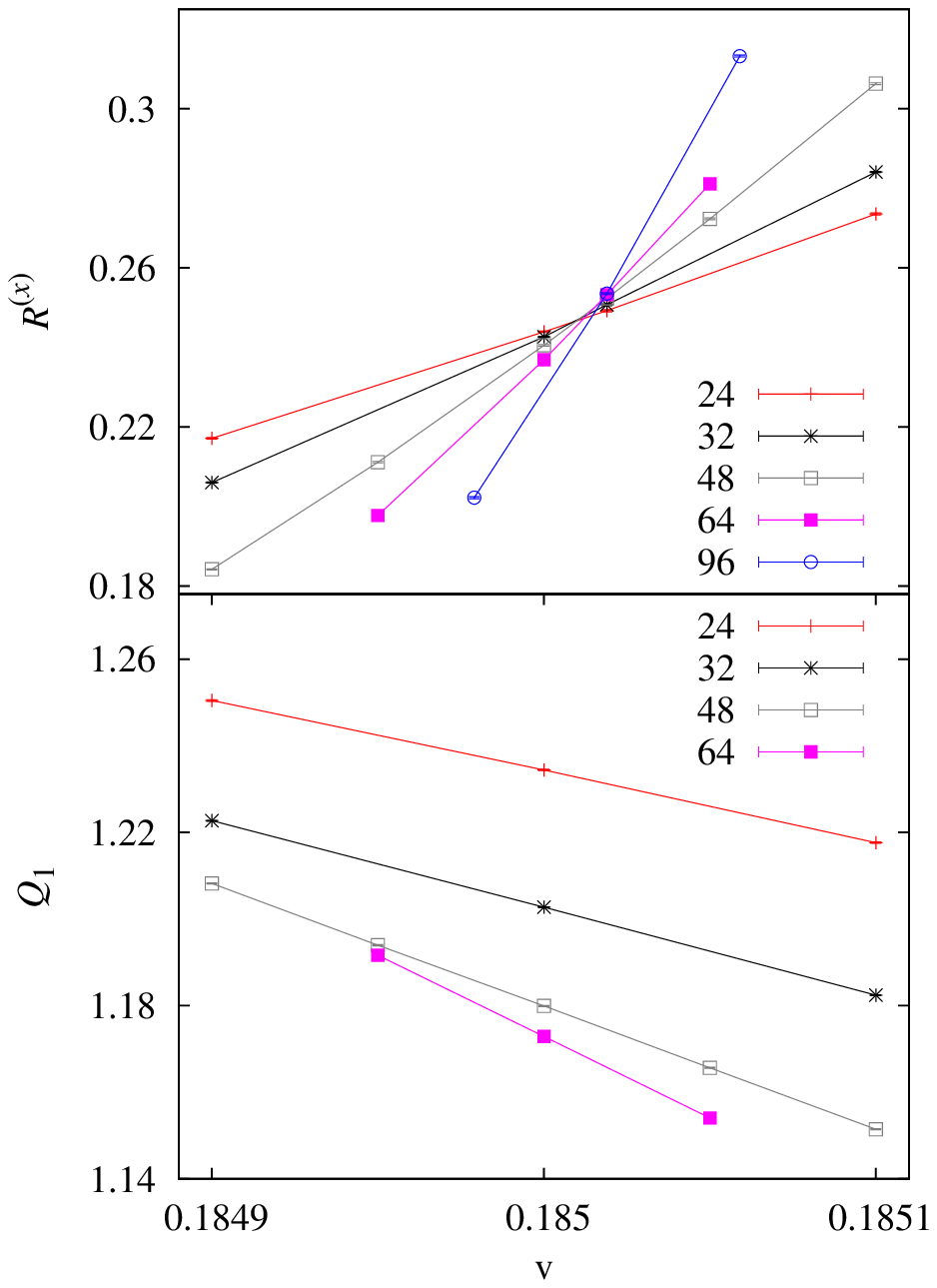}
  \caption{Plots of $R^{(x)}$ and $Q_1$ versus $v$ for the leaf-excluded model on the simple-cubic lattice.}  
 \label{Fig:3DRxQ1}
  \end{figure}

 We again began by fitting the data for $R^{(x)}$, $R^{(d)}$ and $Q_1$, $Q_2$ to the ansatz~\eqref{Fit:fitnearpc} in order to estimate
 $\vc$. Fig.~\ref{Fig:3DRxQ1} plots $R^{(x)}$ and $Q_1$ versus $v$, which again clearly shows that $R^{(x)}$ suffers from only very weak corrections to scaling. In each case of fits, leaving $y_i$ free resulted in unstable fits. Instead, we fixed $y_i = -1.2$, which is numerically
 estimated in~\cite{WangZhouZhangGaroniDeng13} to be the leading correction exponent for three-dimensional percolation. For comparison, we
 performed fits with different combinations of the terms $b_1L^{-1.2}$ or $b_2L^{-2}$ present. The best estimates were obtained from
 $R^{(x)}$, which yield $\vc =0.185\,022(3)$ and $y_t = 1.143(8)$. We also estimated the universal amplitudes $R^{(x)}_c=0.260(4)$ and
 $R^{(d)}_c = 0.083(4)$.
 The universal amplitudes for $Q_1$ and $Q_2$ cannot be precisely estimated due to the strong finite-size corrections.
 
 Simulating at our estimated $\vc$, we then fitted the data for $g^{(x)}_{bR}$, $C_1$, $\chi$ to the ansatz~\eqref{Fit:fitatpc} to estimate
 $y_t$ and $y_h$. Both the two correction terms were included in the fits. The fits of $g^{(x)}_{bR}$ yields $y_t = 1.142(7)$. From $C_1$
 we estimate $y_h = 2.513(5)$. However, we find that it is difficult to estimate $y_h$ from $\chi$ due to the strong finite-size
 corrections. 

  \begin{figure}
  \includegraphics[scale=0.65]{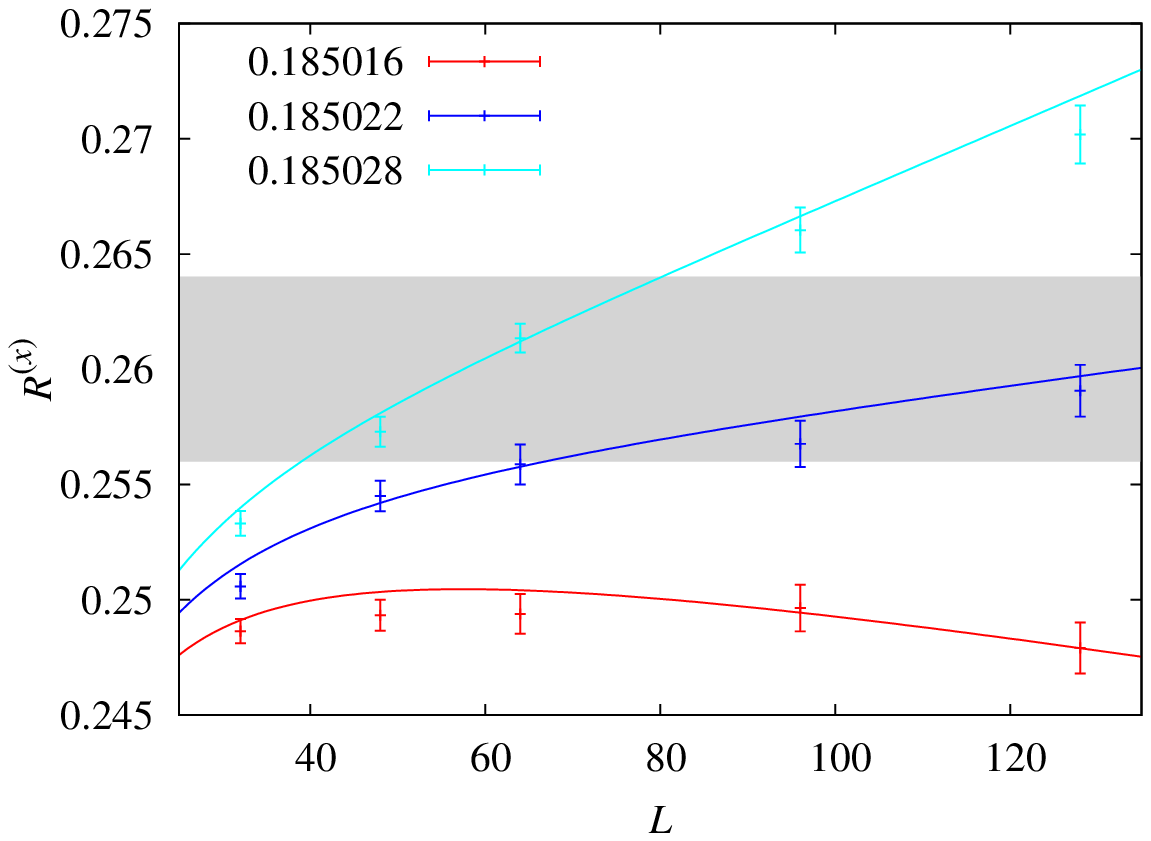}
  \caption{Plots of $R^{(x)}(p,L)$ versus $L$ for fixed values of $p$, for the three-dimensional leaf-excluded model. The curves correspond
    to our preferred fit of the Monte Carlo data. The shaded grey strips indicate an interval of one error bar above and below the estimate
    $R^{(x)}_c=0.260(4)$.}
 \label{Fig:check_pc_3d}
  \end{figure}

\begin{figure}
  \includegraphics[scale=0.8]{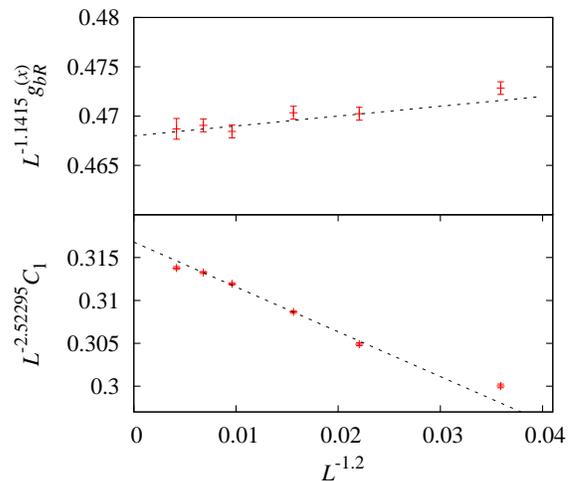}
  \caption{
    Plot of $L^{-1.141\,5}g^{(x)}_{bR}$ and $L^{-2.522\,95}C_1$ versus $L^{-1.2}$.  The values of the critical exponents used on the
    vertical axis correspond to the estimates $y_t=1.1415(15)$ and $y_h=2.522 \,
    95(10)$~\cite{WangZhouZhangGaroniDeng14E,WangZhouZhangGaroniDeng13}. The straight lines are simply to guide the eye.}
 \label{Fig:3Dytyh}
  \end{figure}

 We again illustrate our estimated $\vc$ by plotting $R^{(x)}$ versus $L$ for fixed values of $v$ around our central estimate of $\vc$. The figure confirms that the true value of $\vc$ lies within two error bars of our central estimate. In this
 case however, the curvature suggests the central estimate lies slightly above the true value of $\vc$.  See Fig.~\ref{Fig:check_pc_3d}.
 Our estimates for the critical threshold, critical exponents and wrapping probabilities on the simple-cubic lattice are summarized in
 Tab.~\ref{Tab:Results}. The agreement with the corresponding values for standard three-dimensional
 percolation~\cite{WangZhouZhangGaroniDeng13} strongly suggests the leaf-excluded model is in the percolation universality class. As further
 illustration, Fig.~\ref{Fig:3Dytyh} shows plots of $L^{-y_t}g^{(x)}_{bR}$ and $L^{-y_h}C_1$ versus $L^{y_i}$, using percolation exponent
 values taken from~\cite{WangZhouZhangGaroniDeng13}.
 
\begin{table*}[htbp]
  \begin{tabular}[t]{l|l|lllllll}
    \hline
   $d$& Model     & $\vc$              &  $y_t$        & $y_h$          &  $R_c^{(x)}$   &  $R_c^{(d)}$       & $Q_{1,c}$          &  $Q_{2,c}$    \\
    \hline
   {\multirow{2}{*}{2}}
   & Leaf-excluded          &  0.355\,247\,5(8)  & $0.751(1)$    & $1.895\,8(1)$  &  $0.5212(2)$   & $0.351\,7(1)$     &  $1.041\,46(10)$   & $1.148\,7(2)$ \\
   & Percolation~\cite{Nienhuis84,Pinson94,ZiffLorenzKleban99,HuBloteDeng12}
   			      &  1                 & $3/4$         & $91/48$        &$0.521\,058\,290$ & $0.351\,642\,855$ & $1.041\,48(1)$     & $1.148\,69(3)$ \\
        \hline 
   {\multirow{2}{*}{3}}
   & Leaf-excluded          &  0.185\,022(3)     & $1.143(8)$     & $2.513(5)$     &  0.260(4)        & 0.083(4)          &  -       & -   \\
   & Percolation~\cite{WangZhouZhangGaroniDeng13,WangZhouZhangGaroniDeng14E}
                  &  0.331\,224\,4(1)  & $1.141\,5(15)$ & $2.522\,95(15)$ & 0.257\,80(6) & 0.080\,44(8)      &  1.155\,5(3)       & 1.578\,5(5) \\
   \hline
  \end{tabular}
  \caption{Summary of our estimates for the thresholds $\vc$, critical exponents $y_t$ and $y_h$, and wrapping probabilities for the
    leaf-excluded model. A comparison with standard bond percolation is also included.}
  \label{Tab:Results}
\end{table*}

\section{Discussion}
\label{Discussion}
We have introduced in this paper the leaf-excluded model, and investigated its critical behavior.  Monte Carlo simulations of the
leaf-excluded model were carried out on the square and simple-cubic lattices with periodic boundary conditions.  By studying wrapping
probabilities, we estimated the critical thresholds $\vc = 0.355\,247\,5(8)$ (square) and $\vc = 0.185\,022(3)$ (simple-cubic). The critical
exponents $y_t$ and $y_h$ and wrapping probabilities were found to be consistent with those for standard percolation, which indicates that
the phase transition of the leaf-excluded model belongs to the percolation universality class.
 
As mentioned in the Introduction, rather than enforcing the absence of degree 1 vertices, as we have considered in the current work, one
could more generally forbid any specified set of vertex degrees. A very familiar example is to exclude odd vertices, in which case one
obtains the high-temperature expansion of the Ising model. Dimer, monomer-dimer, and fully-packed loop models also fit into this framework.
A general forbidden-degree model of this kind was studied on the complete graph (i.e. in mean field) from a probabilistic perspective
in~\cite{GrimmettJanson10}, however questions of universality were not considered.  It would be of interest to understand systematically how
the choice of forbidden vertex degrees affects the resulting universality class.
 
Finally, it would be natural to consider a generalization of~\eqref{leaf-excluded distribution} which included a cluster fugacity, in
addition to the bond fugacity. Such a model would correspond to the Fortuin-Kasteleyn model conditioned on the absence of leaves.

\section{Acknowledgments}
\label{Acknowledgments}
This work is supported by the National Nature Science Foundation of  China under Grant No. 11275185, and the Chinese Academy of Sciences. It
was also supported under the Australian Research Council's Discovery Projects funding scheme (project number DP110101141), and T.M.G. is the
recipient of an  Australian Research Council Future  Fellowship (project number FT100100494).  The  simulations were carried out  in part on
NYU's ITS cluster, which is partly supported by NSF Grant No. PHY-0424082.  In addition, this research was undertaken with the assistance of
resources provided  at the NCI  National Facility  through the National  Computational Merit Allocation  Scheme supported by  the Australian
Government.   Y.J.D  also  acknowledges  the  Specialized  Research  Fund   for  the  Doctoral  Program  of  Higher  Education  under  Grant
No. 20113402110040 and  the Fundamental Research Funds for the Central  Universities under Grant No. 2340000034. T.M.G.  is grateful for the
hospitality shown by  the University of Science and  Technology of China at which  this work was completed, particularly  the Hefei National
Laboratory for Physical Sciences at Microscale.


\end{document}